\documentclass[
reprint,
superscriptaddress,
amsmath,
amssymb,
aip,
longbibliography,
jcp,
showpacs,
floatfix
]{revtex4-2}

\usepackage[mathlines]{lineno}
\usepackage{graphicx}
\usepackage[margin=1in]{geometry}
\usepackage{graphicx}
\usepackage[table]{xcolor}
\usepackage{svg}

\usepackage{tabularx}
\usepackage{multirow}
\usepackage{dcolumn}
\newcolumntype{d}[1]{D{.}{.}{#1}}
\newcolumntype{L}[1]{>{\raggedright\arraybackslash}p{#1}}
\newcolumntype{C}[1]{>{\centering\arraybackslash}p{#1}}
\newcolumntype{R}[1]{>{\raggedleft\arraybackslash}p{#1}}
\usepackage{booktabs}

\usepackage{comment}
\usepackage{enumitem}

\usepackage[hidelinks]{hyperref}
\definecolor{color1}{rgb}{0,0.25,0.70}
\hypersetup{colorlinks=true,
    linkcolor={color1},
    citecolor={color1},
    urlcolor={color1}
}

\usepackage{physics}
\usepackage{bm}
\usepackage{siunitx}
\DeclareSIUnit{\au}{a.u.}
\usepackage[version=4]{mhchem}

\usepackage[]{xcolor}
\usepackage{soul}
\usepackage{cancel,siunitx}

\makeatletter
\newsavebox{\@brx}
\newcommand{\llangle}[1][]{\savebox{\@brx}{\(\m@th{#1\langle}\)}%
  \mathopen{\copy\@brx\kern-0.5\wd\@brx\usebox{\@brx}}}
\newcommand{\rrangle}[1][]{\savebox{\@brx}{\(\m@th{#1\rangle}\)}%
  \mathclose{\copy\@brx\kern-0.5\wd\@brx\usebox{\@brx}}}
\makeatother

\DeclareSIUnit\angstrom{\text {Å}}
\DeclareSIUnit\bohr{\text {\ensuremath {a}}_{0}}

\graphicspath{ {./figures/} }

\begin{document}

\preprint{APS/123-QED}

\title{Understanding the polaritonic ground state in cavity quantum electrodynamics}

\author{Tor S. Haugland}
 \affiliation{Department of Chemistry, Norwegian University of Science and Technology, 7491 Trondheim, Norway%
}

\author{John P. Philbin}
 \affiliation{College of Letters and Science, University of California, Los Angeles, CA 90095, USA%
}

\author{Tushar K. Ghosh}
 \affiliation{Department of Chemistry, Purdue University, West Lafayette, IN 47907, USA%
}

\author{Ming Chen}
 \email{chen4116@purdue.edu}
 \affiliation{Department of Chemistry, Purdue University, West Lafayette, IN 47907, USA%
}

\author{Henrik Koch}
 \email{henrik.koch@sns.it}
 \affiliation{Department of Chemistry, Norwegian University of Science and Technology, 7491 Trondheim, Norway}
 \affiliation{Scuola Normale Superiore, Piazza dei Cavalieri, 7, 56124 Pisa, Italy%
}

\author{Prineha Narang}
 \email{prineha@ucla.edu}
 \affiliation{College of Letters and Science, University of California, Los Angeles, CA 90095, USA%
}

\date{\today}

\begin{abstract}
Molecular polaritons arise when molecules interact so strongly with light that they become entangled with each other.
This light-matter hybridization alters the chemical and physical properties of the molecular system and allows chemical reactions to be controlled without the use of external fields.
We investigate the impact of strong light-matter coupling on the electronic structure using perturbative approaches and demonstrate that Rayleigh-Schrödinger perturbation theory can reproduce the ground state energies in optical cavities to comparable accuracy as \textit{ab initio} cavity quantum electrodynamics methodologies for currently relevant coupling strengths.
The method is effective in both low and high cavity frequency regimes and straightforward to implement via response functions.
Furthermore, we establish simple relations between cavity-induced intermolecular forces and van der Waals forces.
These findings provide valuable insight into the manipulation of ground-state polaritonic energy landscapes, shedding light on the systems and conditions in which modifications can be achieved.
\end{abstract}

\maketitle

\section{Introduction}
Cavity quantum electrodynamics (cQED) is the study of systems in optical cavities, where molecules can interact strongly with the standing waves of light and form correlated molecule-photon states known as molecular polaritons.\cite{Flick2018a,Scholes2020}
The simplest example of an optical cavity is a Fabry-Pérot cavity, consisting of two highly reflecting mirrors separated by distances ranging from nanometers to centimeters.\cite{Benz2016,Marissa2023}
The presence of the cavity has been shown to induce significant changes in the properties of molecular systems, such as changes in reactivity\cite{Hutchison2012,Ahn2023} and nonradiative energy transfer.\cite{zhong2017energy}
Other experiments have also shown strong light-matter coupling with single or few molecules in plasmonic nanocavities.\cite{chikkaraddy2016single,Baumberg2022}
Recently, there was also an experiment indicating that intermolecular forces play a role in vibrational strong-coupling.\cite{Maciej2023}
Theoretical studies in polaritonic chemistry have investigated charge transfer,\cite{SchaeferPNAS2019,Avriller2021} orientational effects,\cite{Liberato2017,Keeling2018,philbin2022molecular} relations between local and collective effects,\cite{Sidler2021} changes to chemical reactivity.\cite{mandal2019,Philbin2022,Schafer2022,Joel2023}
For a more complete overview of polaritonic chemistry, several recommended reviews are available.\cite{RibeiroChemSci2018,Li2020,huo_review,ruggenthaler2022}

Molecular polaritons require a quantum description of both the molecular and light degrees of freedom to describe their entanglement.
To accurately describe this entanglement for many different system sizes, many electronic structure methods have been extended to include the cQED effects.\cite{Ruggenthaler2014,Rivera2019,Haugland2020,Flick2020,riso2022MO,Bauer2023}
However, the effects on the electronic ground state are much weaker than the effect of electron correlation considered in quantum chemistry.
Thus, a perturbative approach that is agnostic to the choice of the electronic structure method should be able to accurately describe the effect of light-matter coupling on the electronic ground state.

To this end, herein we derive, implement, and test perturbative expressions for electron-photon interactions in molecular polaritons using response functions.
Within an existing response theory framework for any electronic structure method, coupled cluster theory in our case, we demonstrate how \textit{ab initio} cQED calculations can be performed.
Using this framework, we show how intermolecular pair potentials, screening effects, and orientational effects change when experimental parameters such as the mirror-mirror distance, volume, and concentration are tuned.
The key equations are conceptually simple, offering valuable insights into the polaritonic ground state.

\section{Theory}
The Pauli-Fierz (PF) Hamiltonian describes the total wave function of charged particles and quantized electromagnetic fields (\textit{i.e.} photon modes) and, thus, is relevant for understanding the molecular polaritons that arise when molecules are placed inside optical cavities.
The PF electronic Hamiltonian in the length gauge and long-wavelength approximation for a single mode in atomic units is\cite{ruggenthaler2022}
\begin{equation} \label{eq:hamiltonian}
    H = H_e + \omega_c b^\dagger b + \lambda \sqrt{\frac{\omega_c}{2}} d_\epsilon (b + b^\dagger) + \frac{1}{2} \lambda^2 d_\epsilon^2,
\end{equation}
where $H_e$ is the electronic Hamiltonian, $\omega_c$ is the photon frequency, $\lambda$ is the light-matter coupling strength and $\vec{d}$ is the molecular dipole moment.
We will use subscript $\alpha$, $\beta$, $\gamma$ and $\delta$ to refer to the Cartesian indices $xyz$, and subscript $\epsilon$ to indicate the direction of the transversal polarization vector $\vec \epsilon$ of the cavity photon mode such that $d_\epsilon = \sum_\alpha d_\alpha \epsilon_\alpha$.

In optical cavities, the coupling is given in terms of the effective mode volume $V_{\rm eff}$, $\lambda =  \sqrt{4\pi/V_{\rm eff}}$.
In a Fabry-Pérot cavity, two highly reflective parallel mirrors separated by a distance $L$ form the optical cavity.
The mirror separation distance and speed of light $c$ set the frequency of the standing waves in the cavity via $L = c/2\omega_c$.
Thus, the light-matter coupling strength in a Fabry-Pérot cavity can be written in terms of the frequency and area of the mirrors $A_{\rm eff}$ as 
\begin{equation} \label{eq:fabry_coupling}
    \lambda(\omega_c) = \sqrt{\frac{2\omega_c}{A_{\rm eff}}}.
\end{equation}

\subsection{Perturbation theory}
Perturbation theory has commonly been used to treat electron correlation in physics and chemistry,\cite{HelgakerBook2000} and in the case of intermolecular interactions, the equations are straightforwardly obtained.\cite{Stone2013}
In quantum chemistry, electron correlation can be described by orbital-based perturbation theory, such as Møller-Plesset perturbation theory (MP2) and algebraic diagrammatic construction (ADC), which uses canonical orbitals to describe the electronic states.
Recently, Bauer and Dreuw investigated orbital-based perturbation theory for molecular polaritons using MP2 and ADC(2), and their results compared favorably with QED-CCSD.\cite{Bauer2023}
In this work, we utilize the more general state-based Rayleigh-Schrödinger perturbation theory (RSPT). RSPT uses the eigenstates of the unperturbed system and is identical to the Møller-Plesset perturbation theory when the unperturbed Hamiltonian is the Fock operator.
Expressions in perturbation theory are standardly written as sum-over-states and thus, naively, require the calculation and inclusion of many excited states to be quantitatively accurate.\cite{Wibgerg2006}
The computational cost of calculating many excited states can be mitigated by reformulating the sum-over-states expressions into equivalent linear systems given in terms of response functions.\cite{Jeppe1985,Linderberg2004}

A famous example of the quantum electromagnetic vacuum affecting the ground state of a system is the Lamb shift in free space. 
The shift refers to the 0.03~cm$^{-1}$ (4~$\mu$eV) energy splitting between the 2S$^{1/2}$ and 2P$^{1/2}$ states in the hydrogen atom and arises from the different dipole fluctuations in the two states.
Inside an optical cavity, the Lamb shift is altered as a result of the different boundary conditions compared to free space and stronger coupling of the electromagnetic field to the molecular system.
Light-matter coupling in cavity QED is often denoted as ``strong'', in the sense that the loss rate is smaller than the Rabi period and that it gives a significant change in the excited states.\cite{FriskKockum2019}
In the ground state, the electron-photon correlation effects are much weaker than the correlation between electrons in most molecular polariton systems.
Thus, even in the ``ultra-strong'' regimes where the Rabi frequency is almost as large as the cavity frequency, the modification of the ground state induced by moving a molecule from free space into an optical cavity should be relatively small.
These facts were the basis for our hypothesis that many of the cavity-induced changes to the molecular ground state should be able to be accurately computed using a perturbation theory based method. 

To find the modified ground state of the correlated light-matter system, we start with the bare molecule and electromagnetic field. 
Following Rayleigh-Schrödinger perturbation theory as in Ref.~\citenum{philbin2022molecular}, we expand our wave function and energy perturbatively,\cite{HelgakerBook2000}
\begin{equation}\label{eq:rspt_eq}
    \ket{0} = \sum_{k=0} \ket{0^{(k)}}, \quad E = \sum_{k=0} E^{(k)}.
\end{equation}
The Hamiltonian we are considering is the PF Hamiltonian given in Eq.~\eqref{eq:hamiltonian}, where we regard $\lambda$ as both the light-matter coupling and the perturbation order.
The parametrization in Eq.~\eqref{eq:rspt_eq} and the PF Hamiltonian are inserted into the Schrödinger equation and sorted by orders of perturbation. To second order in $\lambda$, the ground state energy is given by
\begin{align}\label{eq:rspt2}
    E = E_e + \frac{1}{2}\lambda^2 \qty (\expval{\Delta d_\epsilon^2} - \omega_c \sum_{n\neq 0} \frac{d_{\epsilon, 0n} d_{\epsilon, n0}}{\omega_n + \omega_c}), 
\end{align}
where $\Delta d_\epsilon = d_\epsilon - \expval{d_\epsilon}$ is the dipole fluctuation, $\omega_n$ is the electronic excitation energy ($\hbar \omega_n = E_n - E_0$), $\omega_c$ is the cavity photon frequency, and
the transition dipole moments $\mel{n}{d_\epsilon}{m}$ are denoted $d_{\epsilon, nm}$.
The modification to the ground state energy is relatively small as a result of being quadratic in $\lambda$.
In contrast, the changes to the excited states are linear in $\lambda$ when $\omega_c$ is resonant with the excited state.
Sum-over-states expressions such as the one in Eq.~\eqref{eq:rspt2} converge slowly with the number of states.\cite{Wibgerg2006}
To remove the computationally demanding challenge of determining many excited states, one can rewrite a sum-over-states expression in terms of response functions,\cite{Jeppe1985,Linderberg2004}
\begin{align}\label{eq:lr}
    \llangle A; B \rrangle_{\omega} = -\sum_n \qty(\frac{A_{0n}B_{n0}}{\omega_n - \omega} + \frac{B_{0n}A_{n0}}{\omega_n + \omega}).
\end{align}
The response functions for electron-photon interactions are closely related to the frequency-dependent polarizability,
\begin{equation}
    \alpha_{\alpha\beta}(\omega) = - \llangle d_{\alpha}; d_\beta \rrangle_\omega.
\end{equation}
Specifically, the electron-photon response function that is connected to Eq.~\eqref{eq:rspt2} is
\begin{align}\label{eq:ep_int_lr}
    \llangle d_\epsilon b; d_\epsilon b^\dagger \rrangle_{0}
    &= - \frac{1}{\omega_c} \expval{d_\epsilon}^2 - \sum_{n\neq0} \qty(\frac{d_{\epsilon, 0n}d_{\epsilon, n0}}{\omega_n + \omega_c}),
    \\ \nonumber
    &= - \frac{1}{\omega_c}\expval{d_\epsilon}^2 - \alpha^+_{\epsilon\epsilon}(\omega_c),
\end{align}
where we have introduced $\alpha^+$ that is identical to the dipole-dipole response function with the second sum-of-terms given in Eq.~\eqref{eq:lr}.
Inserting $\alpha^+$ into Eq.~\eqref{eq:rspt2}, the expression for the energy becomes
\begin{align}\label{eq:lr2}
    E &= E_e + \frac{1}{2}\lambda^2 \qty (\expval{\Delta d_\epsilon^2} - \omega_c \alpha^+_{\epsilon\epsilon}(\omega_c)).
\end{align}
In the ground state, the bilinear interaction can be thought of as screening the dipole self-energy. As seen from Eq.~\eqref{eq:rspt2}, the sum-over-states term ($\omega_c \alpha^+(\omega_c)$) is always smaller in magnitude than $\expval{\Delta d_\epsilon^2}$ and completely cancels it in the high cavity frequency limit.
Since expectation values and response functions are implemented in most electronic structure packages, Eq.~\eqref{eq:lr2} is relatively easy to implement as it only requires one part of the frequency-dependent polarizability.
In Section~\ref{sec:CC}, we introduce the electron-photon response function in Eq.~\eqref{eq:ep_int_lr} within coupled cluster theory.

\subsection{Intermolecular interactions}
The electronic ground state of molecular polaritons are known to be modified by cavity-induced intermolecular forces.\cite{galego2019cavity,Haugland2021,chemrxiv_huo_gs2023,Milonni2023}
These cavity-induced effects have been included in several molecular dynamics simulations in optical cavities.\cite{cavmd,10.1021/acs.jctc.3c00137,philbin2022molecular}
In this section, we find perturbative expressions for the intermolecular forces in terms of known molecular properties, both for molecules and atoms.

The interaction term for intermolecular forces is the dipole-dipole interaction,\cite{Stone2013} given here in terms of the intermolecular interaction tensor $\mathcal{T}_{\alpha\beta}$ and the dipole moment of molecules $A$ and $B$,
\begin{align}
    V^{AB}_{\rm dip} &= -\sum_{\alpha\beta} d^A_{\alpha} \mathcal{T}_{\alpha\beta} d^B_{\beta}
    \\
    \mathcal{T}_{\alpha\beta} &= -\frac{1}{R^3} \qty(\delta_{\alpha\beta} - \frac{3 R_\alpha R_\beta}{R^2}).
\end{align}
The PF Hamiltonian for describing two molecules $A$ and $B$ inside an optical cavity is then approximated by
\begin{equation}
    H^{AB} = H^A + H^B + V^{AB}_{\rm dip} + \lambda^2 d^A_\epsilon d^B_\epsilon,
\end{equation}
where $H^A$ and $H^B$ refer to the PF Hamiltonian for the individual molecules, Eq.~\eqref{eq:hamiltonian}.

The intermolecular interactions in free space lead to the well-known dipole-dipole interactions and the van der Waals forces.
In this section, we restrict our derivations to non-polar molecules.
The van der Waals forces show up in second-order perturbation theory with the dipole-dipole interaction,
\begin{align} \label{eq:vr_abab}
    E^{AB}_{\rm vdW} &= \frac{1}{2} \llangle V_{\rm dip}^{AB}; V_{\rm dip}^{AB} \rrangle_0.
\end{align}
To separate $A$ and $B$, we use the following relation between the frequency-dependent polarizability and the dipole response functions (see Appendix~\ref{sec:casimir_polder}),
\begin{align}\label{eq:vabvab_lr}
    \llangle d^A_{\alpha} d^B_{\gamma}; d^A_{\beta} d^B_{\delta} \rrangle_0 &= - \frac{1}{\pi} \alpha^{AB}_{\alpha\beta\gamma\delta},
\end{align}
where we introduced the integral over the frequency-dependent polarizability,
\begin{align}
    \alpha^{AB}_{\alpha\beta\gamma\delta} = \int_0^\infty \alpha^A_{\alpha\beta}(i\omega) \alpha^B_{\gamma\delta}(i\omega) \dd{\omega}.
\end{align}
These types of integrals can be evaluated using Cauchy moments and Padé approximants as described in Refs.~\citenum{suleman1979pade} and~\citenum{10.1063/1.474223}.
Inserting this relation into the van der Waals energy expression of Eq.~\eqref{eq:vr_abab}, we obtain
\begin{align}\label{eq:casimirpolder}
    E^{AB}_{\rm vdW} &= -\frac{1}{2\pi} \sum_{\alpha\beta\gamma\delta} \mathcal{T}_{\alpha\gamma} \mathcal{T}_{\beta\delta} \alpha^{AB}_{\alpha\beta\gamma\delta}.
\end{align}

Similarly to how the standard van der Waals force is derived, we find a cavity-induced van der Waals force to be
\begin{align} \label{eq:cvdw_force}
    E^{AB}_{\rm cvdW} &= \llangle \lambda^2 d^A_\epsilon d^B_\epsilon; V^{AB}_{\rm dip} \rrangle_0 
    \\ \nonumber
    &= \frac{\lambda^2}{\pi} \sum_{\alpha\beta} \mathcal{T}_{\alpha\beta} \alpha^{AB}_{\epsilon\alpha\beta\epsilon}.
\end{align}
We have disregarded the second-order term that arises from the bilinear interaction that enters the expression via the quadratic response functions. We expect this term to be small in the low-frequency regime and only have a screening effect similar to the one-body energy in Eq.~\eqref{eq:rspt2}.
Note that the distance dependence for the cavity-induced van der Waals force is $R^{-3}$ unlike the $R^{-6}$ of standard van der Waals forces.
This $R^{-3}$ dependence is similar to the $R^{-3}$ dependence seen for molecules in external electric fields and the dipole-dipole interactions.\cite{Milonni1996}

The first cavity-induced interaction which is independent of distance (i.e. a collective effect) is a fourth-order perturbation contribution. For the dipole self-energy, this energy modification is
\begin{align}\label{eq:E_long_lr_lambda}
    E_{\rm long}^{AB} &= \frac{1}{2} \llangle \lambda^2 d^A_\epsilon d^B_\epsilon; \lambda^2 d^A_\epsilon d^B_\epsilon \rrangle
    \\ \nonumber
    &= - \frac{\lambda^4}{2\pi} \alpha^{AB}_{\epsilon\epsilon\epsilon\epsilon}.
\end{align}
There are also equivalent fourth-order terms that arise from a mixed bilinear/self-energy interaction and from 
a purely bilinear interaction that require the evaluation of quadratic and cubic response functions, respectively.
We disregard these other terms as we expect them to only screen the effects of the dipole self-energy.
Because Eq.~\eqref{eq:E_long_lr_lambda} is independent of the intermolecular distance $R$, it depends on the number of molecules within the cavity.
Thus, the higher body terms become more and more important with the number of molecules in the cavity.
The $n$-body long range interaction can be evaluated similarly to $n$-body van der Waals interactions, and is given by the integral
\begin{equation}\label{eq:n_body_z}
    E_{\rm long}^{AB\dots Z} = \frac{(-1)^{n-1}}{\pi} \int_0^\infty \alpha^A_{\epsilon\epsilon}(i\omega) \alpha^B_{\epsilon\epsilon}(i\omega) \dots \alpha_{\epsilon\epsilon}^Z(i\omega) \dd{\omega}
\end{equation}
This integral can also be used to evaluate the $1$-body energy, \textit{i.e.} the dipole self-energy.
Thus, if one has an accurate expression for the frequency-dependent polarizability, for instance by using the Cauchy moments, one can evaluate the $n$-body interactions accurately.

To summarize, we derived expressions for the one-body energy, Eq.~\eqref{eq:lr2}, and expressions for the two-body interactions, Eqs.~\eqref{eq:casimirpolder}, \eqref{eq:cvdw_force} and \eqref{eq:E_long_lr_lambda} in a strong light-matter coupling environment. Ignoring higher than two-body interactions, the total ground state energy is
\begin{align}\label{eq:total_en}
    E_{\rm tot} = \sum_A E^A + \frac{1}{2} \sum_{A\neq B} \qty(E^{AB}_{\rm vdW} + E^{AB}_{\rm cvdW} + E^{AB}_{\rm long}).
\end{align}
Usually, three-body effects are relatively small, but there are cases such as argon gas where they are important.\cite{Cencek2013} 
However, because the higher-body effects in cavities do not decrease with distance, they can be important in every type of molecular system.
Especially in cases of strong coupling and a large number of molecules.
In the following section, we show the energy of long-range higher-body effects using an approximate relation for the polarizability.

\subsubsection{Atoms and isotropic molecules}

This general expression for the van der Waals interaction energy can be simplified to the well-known $C_6$ expression in the special case where the molecules are isotropic ($\alpha_{\alpha\beta}=\alpha_{\alpha\alpha} \delta_{\alpha\beta}$) given by
\begin{align} \label{eq:vdW}
    E^{AB}_{\rm vdW} &= - \frac{C_6}{R^6},
\end{align}
where we can identify the coefficient for London dispersion forces, $C_6$, as
\begin{align}\label{eq:C6}
    C_6 &= \frac{3}{\pi} \alpha^{AB}_{\alpha\alpha\alpha\alpha}.
\end{align}
The cavity-induced forces also become quite similar to the van der Waals interaction in Eq.~\eqref{eq:vdW},
\begin{align}\label{eq:iso_cvdw}
    E_{\rm cvdW}(R) &= \frac{\lambda^2 C_6}{R^3} \qty(\cos^2(\theta) - \frac{1}{3}),
\end{align}
where $\theta$ is the angle between $\vec{\epsilon}$ and $\vec{R}$.

In molecular systems, most molecules are far apart from each other and only interact with other molecules that are close by ($<\SI{5}{\angstrom}$).
However, in cases of strong-light matter coupling, molecules can interact via the light even though they are separated by a long distance, and the interaction between all molecules must be taken into account.
The energy of $N$ isotropic and identical molecules that are spatially separated can be found by solving the integral Eq.~\eqref{eq:n_body_z}.
Using the approximate polarizability
\begin{equation}
    \alpha(i\omega) = \alpha(0) \qty(\frac{I_\epsilon^2}{I_\epsilon^2 + \omega^2}),
\end{equation}
where $I_\epsilon$ is a parameter related to the mean excitation energy, the cavity-induced energy is found to be
\begin{align} \label{eq:long}
    E_{\rm cav} &= -2I_\epsilon \sum^N_{n=1} \qty(-\frac{\alpha\lambda^{2}}{4})^n {2n - 2 \choose n - 1} {N \choose n}.
\end{align}
See Appendix~\ref{sec:n-body} for the derivation.
As the number of molecules in a cavity increase, so does the change in the total energy.
This is especially important for non-isotropic molecules, where there will be orientational forces because $\alpha_{\epsilon\epsilon}$ will vary with the orientation of the molecules. However, in the case where only the concentration $C$ is kept fixed, $\lambda^2 = 4\pi C / N$, the cavity-induced effects disappear, even when accounting for the many $n$-body interactions.

\subsection{Coupled cluster theory}\label{sec:CC}
Coupled cluster theory employs an exponential parametrization of the wave function. The coupled cluster ansatz is given by
\begin{equation}
    \ket{\rm CC} = e^{T} \ket{\rm HF},
\end{equation}
where the cluster operator $T$ is defined as 
\begin{equation}
    T = \sum_\mu t_\mu \hat \tau_\mu,
\end{equation}
where $t_\mu$ are amplitudes to be determined and $\hat \tau_\mu$ creates an excited determinant $\ket{\mu}$.
This ansatz is size-extensive, ensuring that the energy of isolated molecules are additive, even when $T$ is truncated to not include the full configuration space.
The typical truncation of the wave function is at the single and double excitation level (CCSD), allowing up to double excitations in $\ket{\mu}$.
To evaluate expectation values in coupled cluster theory, one also needs the left state $\bra{\Lambda}$,

\begin{equation}
    \bra{\Lambda} = \bra{\rm HF}(1 + \sum_\mu \bar t_\mu \tau^\dagger_\mu)e^{-T}
\end{equation}
which requires the multipliers $\bar t_\mu$.
The methods to find the amplitudes $t_\mu$ and multipliers $\bar t_\mu$ are described in Ref.~\citenum{HelgakerBook2000}.
The coupled cluster ground state energy is obtained from

\begin{align}
    E_{\rm CC} = \bra{\Lambda} H \ket{\rm CC}.
\end{align}

The QED coupled cluster framework is an extension to standard coupled cluster theory where we allow the excitation space to include both excited determinants and photon number states.\cite{Haugland2020,deprince2020cavitymodulated} The corresponding cluster operator is given in terms of the electronic excitation operator $\tau_\mu$ and the boson creation operator $b^\dagger$,

\begin{equation}
T_{\rm QED} = \sum_{\mu n} t_{\mu n} \tau_{\mu} (b^\dagger)^n.
\end{equation}

In Section \ref{sec:linear_response_derivation} we outline the derivation of the electron-photon response function in coupled cluster theory, which is found to be

\begin{align}
    \llangle d_\epsilon b; d_\epsilon b^\dagger \rrangle_0 &=
    - \frac{1}{\omega_c} \mel{\Lambda}{d_\epsilon}{\rm CC}^2
    \\ \nonumber
    & + \sum_{\mu} \bra{\Lambda} [d_\epsilon, \tau_\mu] \ket{\rm CC} X^{d_\epsilon}_{\mu}(-\omega_c),
\end{align}
where $X^{d_\epsilon}_\mu(\omega)$ is the response vector from the dipole moment operator along the polarization direction.
This expression for the response function can be used together with Eq.~\eqref{eq:lr2} to find the ground state energy of the polaritonic system.
For more linear response theory in polaritonic systems, see Ref.~\citenum{castagnola2023polaritonic}.

\section{Results and Discussion}
To demonstrate the validity of the perturbation theory, we compare the energy of several systems using perturbative (PT) and self-consistent QED calculations. All calculations are performed using the eT program.\cite{eTpaper} The perturbative calculations are performed using HF and CCSD, while the self-consistent calculations are performed with QED-HF and QED-CCSD-12-SD1. The QED-CCSD-12-SD1 is a QED-CCSD method with singles and doubles in the electron and electron one-photon spaces, and up to two photons in the photon space, see Ref.~\citenum{philbin2022molecular}. The extra photon in the photon space is especially important for ground state calculations such as for intermolecular interactions. All calculations are performed using a single photon mode.

\subsection{Single molecule}
\begin{figure*}
    \centering
    \includegraphics[width=0.99\textwidth]{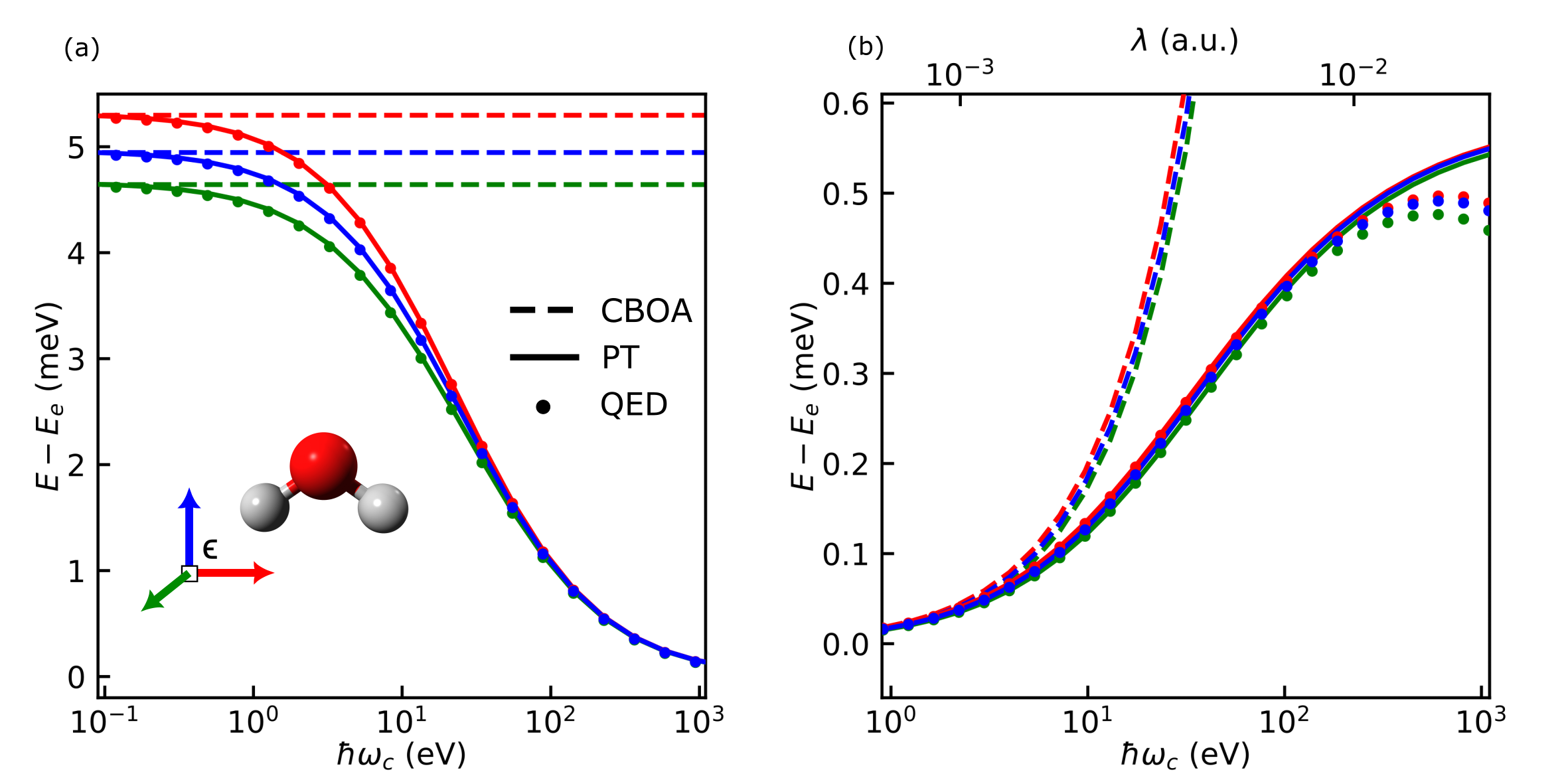}
    \caption{Cavity induced energy for \ce{H2O} using CCSD/aug-cc-pVDZ at different cavity energies. (a) The coupling strength is $\lambda=0.01$ a.u. (b) The coupling strength is $\lambda=0.0152$ a.u. at $\omega_c= 630$ eV, the same as in a cubic \SI{2}{\nm} Fabry-Pérot cavity.}
    \label{fig:h2o_dispersion}
\end{figure*}

We begin by testing the perturbation theory expressions on a single molecule using Eq.~\eqref{eq:lr2}.
In Fig.~\ref{fig:h2o_dispersion}(a), we show a dispersion of the energy, comparing perturbation theory with self-consistent QED for a system of \ce{H2O} in an optical cavity. The effective volume ($V_{\rm eff}$) of the cavity is kept fixed while varying the cavity frequency.

In the high-energy regime, the polaritonic energy is identical to the molecular energy, with negligible contributions from the optical cavity.
This is because the cavity energy levels are too far away from the electronic ones, and thus it follows that the interaction between the molecule and cavity is weak.
When the cavity energy is close to electronic excitation energies, within a couple of \si{\eV}, the interaction between the optical cavity and the molecule increases. This increase in energy is a result of dipole moment fluctuations interacting with the cavity mode.

As the cavity energy is further reduced down to the molecule's vibrational regime, $<\SI{0.1}{\eV}$, the energy curve flattens out.
In this regime, the photon state is ``slow'' compared to the movement of electrons, and we can introduce the cavity Born-Oppenheimer approximation (CBOA).\cite{10.1021/acs.jctc.3c00137}
In the CBOA, the cavity mode is a fixed parameter in the Pauli-Fierz Hamiltonian, just like the nuclei positions in the electronic Hamiltonian, and consequently all electron-photon correlation is ignored.
The ground state energy of the system in the CBOA is given by
\begin{equation} \label{eq:cboa}
    E_{\rm CBOA} = E_{e} + \frac{1}{2} \lambda^2 \expval{\Delta d^2_\epsilon}.
\end{equation}
As shown in Fig.~\ref{fig:h2o_dispersion}, the CBOA becomes exact when the cavity frequency is small.
Therefore, it is only in the frequency regime around $1-\SI{100}{\eV}$ that electron-photon correlation from the bilinear electron-photon interaction is needed.
Both perturbative CCSD, Eq.~\eqref{eq:lr2}, and QED--CCSD can provide the necessary electron-photon correlation in this regime.

The energy regime where electron-photon correlation is important can be more intuitively understood by introducing the \textit{mean electronic excitation energy} $I_\epsilon$.
We obtain an approximate expression for Eq.~\eqref{eq:rspt2} by introducing $\omega_n\rightarrow I_\epsilon$,
\begin{equation} \label{eq:onemol_fit}
    E \approx E_e + \frac{1}{2} \lambda^2 \expval{\Delta d_\epsilon^2} \qty(1 - \frac{\omega_c}{I_\epsilon + \omega_c}).
\end{equation}
From this equation, it is clear that the dipole self-energy is the only contribution at low frequencies, as we already saw in CBOA, Eq.~\eqref{eq:cboa}.
It is also evident from Eq.~\ref{eq:onemol_fit} that in the large cavity frequency limit the bilinear energy and DSE contributions cancel and the system energy returns to the normal electronic energy.
The overall effect of the electron-photon correlation on the ground state can be summarized as a screening effect on the dipole self-energy, the larger the frequency, the larger the screening.

In Fabry-Pérot cavities, the coupling depends on the square root of the frequency, $\lambda \propto \sqrt{\omega}$, see Eq.~\eqref{eq:fabry_coupling}. Thus, the coupling also becomes very strong in in the high frequency regime. The energy dispersion in a Fabry-Pérot cavity is shown in Fig.~\ref{fig:h2o_dispersion}(b).
For cavities that have excitation energies above $100$~eV, the perturbation theory starts to differentiate itself from QED-CCSD.
In particular, the perturbative energy does not decrease with the frequency as the self-consistent QED energy does and instead seems to converge towards a maximum (Fig.~\ref{fig:h2o_dispersion}(b)).
For a single mode Fabry-Pérot cavity, the equivalent cavity length to $100$~eV would be $12$~nm.
For larger cavities than this, the perturbative approach gives a sufficient description of the energy.

We note that in Eq.~\eqref{eq:onemol_fit}, each mode contributes independently.
Given that the coupling is fixed, the energy contribution of modes with higher energy becomes smaller and smaller, unlike in CBOA, where each mode has an energy contribution independent of the mode energy. 
For higher frequencies, we also expect that the effective coupling will be smaller than what we use in both self-consistent and perturbative approaches due to the dipole approximation.
To conclude, we find that perturbation theory accurately reproduces \textit{ab initio} ground state energies for a small molecule in experimentally relevant light-matter coupling strengths today ($\lambda \leq 0.01$~a.u.) and in all cavity frequency regimes.

\subsection{Multiple molecules}
\begin{figure}
    \centering
    \includegraphics[width=0.5\textwidth]{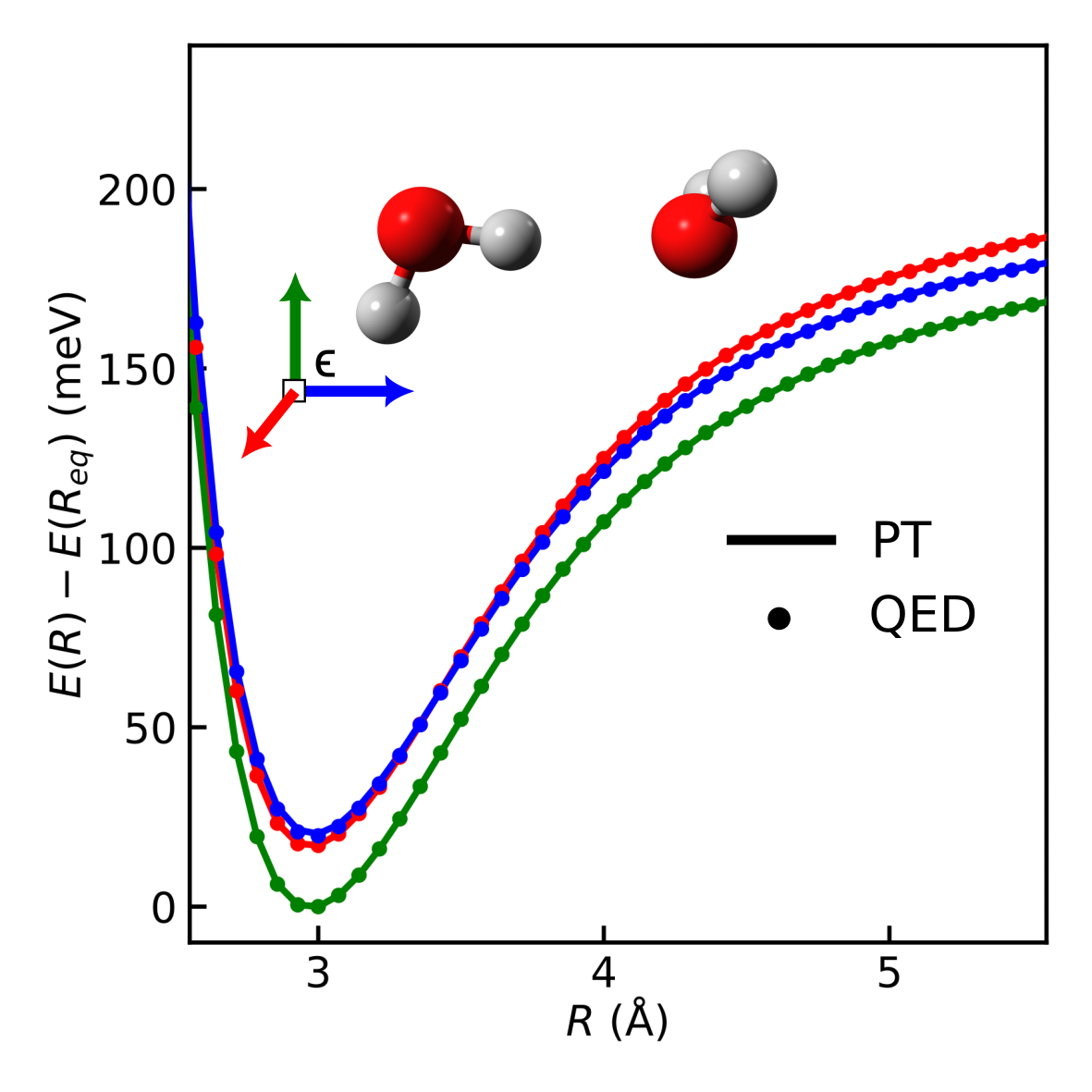}
    \caption{Potential energy surface of a water dimer using CCSD/aug-cc-pVDZ. The coupling is $\lambda=0.05$ a.u. and the cavity energy is $\hbar\omega_c = \SI{2.7}{\eV}$. The energy is relative to the equilibrium geometry.}
    \label{fig:h2o_2_pes}
\end{figure}

A very interesting case is the study of how molecules interact and their collective effects in the polaritonic ground state.
Figure~\ref{fig:h2o_2_pes} shows the ground state potential energy surface of a water dimer inside and outside an optical cavity. We find that both perturbative and self-consistent QED calculations give sufficient descriptions of the energy surface, being only a few percent different for a very strong coupling of $\lambda = 0.05$~a.u. and the agreement improving as $\lambda$ decreases.

From Fig.~\ref{fig:h2o_2_pes}, it is evident that the potential energy surfaces have changed because of the cavity, and therefore that the intermolecular forces must be different. To determine the van der Waals forces fully to $\lambda^2$, one would need to evaluate quadratic response functions. In this section, we limit ourselves to the low cavity frequency regime where screening effects due to the bilinear coupling term are small and the dipole-self energy gives the most significant contributions.

\begin{figure*}
    \centering    
    \includegraphics[width=0.99\textwidth]{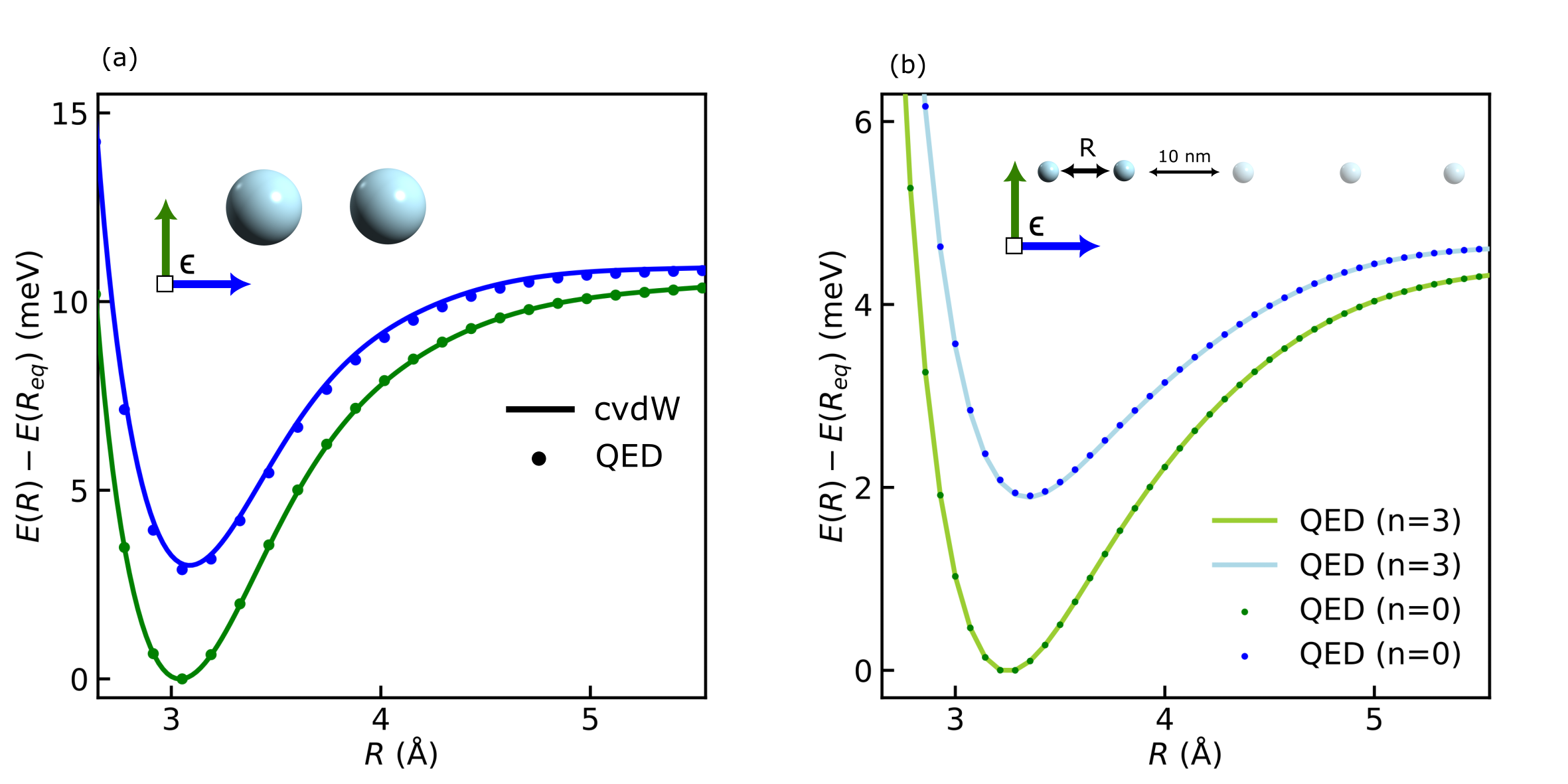}
    \caption{Potential energy curve of (a) \ce{Ne2} using CCSD/d-aug-cc-pVDZ, (b) \ce{Ne2} and \ce{Ne2 + 3 Ne} using CCSD/aug-cc-pVDZ where the 3 \ce{Ne} are spatially isolated by $\SI{10}{\nm}$. In both (a) and (b), the coupling is $\lambda=\SI{0.058}{\au}$ ($V_{\rm eff} = \SI{0.55}{\nm^3}$) and the cavity energy is $\hbar\omega_c = \SI{0.027}{\eV}$.}
    \label{fig:ne_2_pes}
\end{figure*}

For an isotropic case, we choose \ce{Ne} as our test system.
In Fig.~\ref{fig:ne_2_pes}(a) we show the potential energy curve of \ce{Ne2} in d-aug-cc-pVDZ using the $C_6$ parameter from the CCSD/q-aug-cc-pV5Z calculation done in Ref.~\citenum{10.1063/1.474223} together with Eq.~\eqref{eq:iso_cvdw} and the 1-body dipole self-energy. We find that this simple expression reproduces the potential energy surface with high accuracy!

Figure~\ref{fig:ne_2_pes}(b) shows potential energy surfaces of \ce{Ne2} and \ce{Ne2 + 3Ne} in an optical cavity. 
These potential energy surfaces demonstrate that the cavity-induced van der Waals force is only dependent on the distance between the molecules, not on the number of molecules.
Only by decreasing the effective volume and raising the light-matter coupling $\lambda$ can we observe the cavity-induced van der Waals forces.
This makes these interactions very difficult to observe and verify experimentally, as most experiments observe effects that scale with concentration and coupling, not only the coupling.\cite{Marissa2023}
In contrast, for excited states, the energetic splitting between the upper and lower polaritonic states, the Rabi splitting, is proportional to the square root of the number of molecules.

Comparing Fig.~\ref{fig:ne_2_pes}(a) and Fig.~\ref{fig:ne_2_pes}(b), we also see the importance of the basis set on the intermolecular forces. The binding energy with the basis d-aug-cc-pVDZ is twice as deep as the one with basis aug-cc-pVDZ, showing that the diffuse functions play an important role in \ce{Ne} and the determination of ground state effects in cavities.

One consequence of cavity-modified molecular ground states that may be possible to observe experimentally are orientational effects.
Although the cavity-induced van der Waals forces are not collective, the $n$-body long-range orientational effects are, see Eq.~\eqref{eq:long}.
The long-range interactions described by Eq.~\eqref{eq:E_long_lr_lambda} depend on the orientations of both molecule $A$ and molecule $B$.
The polarizability along the polarization of the field is minimized along the same direction for every molecule.
In the case of identical molecules, all molecules will become parallel so that the one-body energy in Eq.~\eqref{eq:lr2} is minimal. However, the two body effects in Eq.~\eqref{eq:long} is maximized when the maximum polarizability is along the polarization direction. Thus, as more and more molecules are put inside the cavity, the orientational effects decrease. 
These orientational effects are shown using path integral molecular dynamics simulations by us in Ref.~\citenum{philbin2022molecular}.
The orientational effects will disappear for large volumes, even if the coupling and concentration stay large.
This is because the force per molecule becomes increasingly small as the cavity volume increases.
Additionally, in our non-relativistic Pauli-Fierz Hamiltonian, retardation effects are not taken into account.
For distances where retardation effects start to become significant, around $\SI{137}{\bohr} = \SI{7.3}{\nm}$, we expect that the orientational effects will decrease, similar to how van der Waals forces become $R^{-7}$ instead of the usual $R^{-6}$.
Thus, the orientational effects are likely to only be observable in low-concentration nanometer-sized systems.
An experiment showing that orientational effects decrease with concentration in a fixed cavity volume will demonstrate that collective effects also play a role in the ground state.

\section{Conclusions}
We have introduced a perturbative approach to determine the ground state of molecules strongly interacting with light.
We demonstrated the accuracy of this perturbative approach within a coupled cluster framework, but this perturbative approach can be implemented using the response functions from any electronic structure theory.
This formalism makes it clear that the changes in the ground state energy are closely related to frequency-dependent polarizabilities and that the cavity-induced intermolecular forces can be linked to the van der Waals forces via the $C_6$ coefficient.
In contrast to polaritonic excited states, collective effects decrease cavity-induced effects in the ground state.

The potentials for one-, two-, and $n$-body interactions introduced in this work can be used in molecular dynamics simulations and serve as tools to understand and investigate changes in chemical and physical properties in strongly coupled light-matter systems.
The perturbative methodology is consistent with self-consistent QED, elucidating that the impact of strong light-matter coupling values on the ground state fall within the weak perturbative regime.
Thus, we think that perturbation theory should be considered as a viable alternative to self-consistent QED when investigating the polaritonic ground state moving forward.

\begin{acknowledgments}
We thank Sara Angelico for helpful discussions about the cavity Born-Oppenheimer approximation. This work is supported by the Department of Energy QIS Program Grant Number DE-SC0022277. T.S.H. and H.K. acknowledge funding from the Research Council of Norway through FRINATEK project 275506 and the European Research Council under the European Union’s Horizon 2020 Research and Innovation Programme grant agreement No. 101020016. T.K.G. and M.C. acknowledge support from Purdue startup funding. This work is supported by the US Department of Energy's 2023 Innovative and Novel Computational Impact on Theory and Experiment (INCITE) award at the Oak Ridge Leadership Computing Facility (OLCF) which is a DOE Office of Science User Facility supported under Contract DE-AC05-00OR22725. This research also used resources of the National Energy Research Scientific Computing Center, a DOE Office of Science User Facility supported by the Office of Science of the U.S. Department of Energy under Contract No. DE-AC02-05CH11231 using NERSC award BES-ERCAP0025026. P.N. is a Moore Inventor Fellow and gratefully acknowledges support through Grant GBMF8048 from the Gordon and Betty Moore Foundation.
\end{acknowledgments}

\section{Appendix}
\subsection{van der Waals energy} \label{sec:casimir_polder}
The dipole-dipole interaction to second order perturbation theory is found to be
\begin{align} \label{eq:ap_vdw}
    E^{AB}_{\rm vdW} &= \frac{1}{2} \llangle V^{AB}_{\rm dip}; V^{AB}_{\rm dip} \rrangle
    \\ \nonumber
    &= \frac{1}{2} \sum_{\alpha\beta\gamma\delta} \mathcal{T}_{\alpha\gamma} \mathcal{T}_{\beta\delta} \llangle d^A_{\alpha} d^B_{\gamma}; d^A_{\beta} d^B_{\delta} \rrangle_0 
\end{align}
From the definition of linear response we find that the response function in Eq.~\eqref{eq:ap_vdw} is
\begin{align}\label{eq:ap_vabvab_lr}
    \llangle d^A_{\alpha} d^B_{\gamma}; d^A_{\beta} d^B_{\delta} \rrangle_0 &= -2 \sum_{\alpha\beta\gamma\delta} \sum_{nm\neq 0}
    \frac{d_{n0,\alpha}^Ad_{n0,\beta}^A d_{m0,\gamma}^Bd_{m0,\delta}^B}{\omega^A_{n} + \omega^B_{m}}
\end{align}
The denominator can be split up from a sum of two variables into a product of them by introducing the integral
\begin{align}\label{eq:ap_ab_identity}
    \frac{1}{a+b} =  \frac{1}{2\pi} \int_0^\infty \qty(\frac{2a}{a^2+\omega^2})\qty(\frac{2b}{b^2+\omega^2})\dd{\omega}
\end{align}
Using the definition of the frequency dependent polarizability we can find the imaginary frequency dependent polarizability,
\begin{equation}
    \alpha_{\alpha\beta}(i\omega) = \sum_{n\neq0} d_{\alpha,0n}d_{\beta,n0} \qty(\frac{ 2\omega_n}{\omega_n^2 + \omega^2}),
\end{equation}
and inserting these relations into Eq.~\eqref{eq:ap_vabvab_lr} we find the the response function,
\begin{align}\label{eq:ap_vabvab_lr2}
    \llangle d^A_{\alpha} d^B_{\gamma}; d^A_{\beta} d^B_{\delta} \rrangle_0 &= - \frac{1}{\pi} \int_0^\infty \alpha^A_{\alpha\beta}(i\omega) \alpha^B_{\gamma\delta}(i\omega) \dd{\omega}.
\end{align}
Thus, the final expression for van der Waals energy is 
\begin{align}\label{eq:ap_evdw}
    E^{AB}_{\rm vdW} &= -\frac{1}{2\pi}\sum_{\alpha\beta\gamma\delta} \mathcal{T}_{\alpha\gamma} \mathcal{T}_{\beta\delta} \int_0^\infty \alpha^A_{\alpha\beta}(i\omega) \alpha^B_{\gamma\delta}(i\omega) \dd{\omega}.
\end{align}

\subsection{n-body interaction} \label{sec:n-body}
The n-body interaction for atoms or isotropic molecules is given by
\begin{align}
    E^{(n)} &= \frac{1}{\pi} (-1)^{n-1} \lambda^{2n} \int_0^\infty \prod_X^{n} \alpha^X(i\omega)_{\epsilon\epsilon} \dd{\omega}
\end{align}
An approximate relation for the $n$-body long range interaction is found by assuming
\begin{equation}
    \alpha(i\omega) = \alpha I^2 / (I^2 + \omega^2).
\end{equation}
For the energy we find
\begin{align}
    E^{(n)}&= \frac{1}{\pi} (-1)^{n-1} \lambda^{2n}\int_0^\infty \qty(\frac{\alpha I^2}{I^2 + \omega^2})^n \dd{\omega}
    \nonumber \\ \nonumber
    &= \frac{I}{\pi} (-1)^{n-1}\lambda^{2n}\alpha^n \int_0^\infty \qty(\frac{1}{1 + x^2})^n \dd{x}
    \\ \nonumber
    &= \frac{I}{\pi} (-1)^{n-1}\lambda^{2n}\alpha^n \frac{\pi}{2^{2n-1}} {2n - 1 \choose n - 1}
    \\
    &= -2I\qty(-\frac{\alpha\lambda^{2}}{4})^n {2n - 2 \choose n - 1}
\end{align}
The total energy of $N$ spatially separated atoms would be
\begin{align}
    E &= -2I \sum^N_{n=1} \qty(-\frac{\alpha\lambda^{2}}{4})^n {2n - 2 \choose n - 1} {N_{\rm mol} \choose n}
\end{align}
Instead, if the concentration $C$ is fixed, $\lambda^2 = 4\pi C / N$,
\begin{align}
    E &= -2I \sum^N_{n=1} \qty(-\frac{\pi \alpha C}{N})^n {2n - 2 \choose n - 1} {N_{\rm mol} \choose n}
\end{align}

\subsection{CC electron-photon response} \label{sec:linear_response_derivation}
The coupled cluster linear response function for two operators $C$ and $D$ are\cite{10.1063/1.473814}

\begin{align}
     \llangle C; D\rrangle_{\omega} &= \sum_\mu \bra{\Lambda} [C, \tau_\mu] \ket{\rm CC} X_\mu^D(\omega)
     \\ \nonumber &
     + \sum_\mu \bra{\Lambda} [D, \tau_\mu] \ket{\rm CC} X_\mu^C(-\omega)
     \\ \nonumber &
     + \sum_{\mu\nu} F_{\mu\nu} X^D_\nu(\omega) X^C_\mu(-\omega).
\end{align}
Here the response vector $X^C_\mu (\omega)$ is

\begin{equation}
    X^C_\mu (\omega) = - \sum_\nu (\mathbf{A} - \omega \mathbf{I})^{-1}_{\mu\nu} \bra{\nu} e^{-T} C \ket{\rm CC},
\end{equation}
where $\mathbf{A}$ is the coupled-cluster Jacobian, 

\begin{align}
    A_{\mu\nu} &= \bra{\mu}e^{-T} [H, \tau_\nu] \ket{\rm CC},
\end{align}
and $F_{\mu\nu}$ is the so-called F-matrix,

\begin{align}
    F_{\mu\nu} &= \bra{\Lambda} [[H, \tau_\mu], \tau_\nu] \ket{\rm CC}.
\end{align}
To find the electronic-photon response function we insert our electron-photon operators and sum over all electronic and photonic excited states,

\begin{align}
    \llangle d_\epsilon(b+ & b^\dagger); d_\epsilon(b+b^\dagger) \rrangle_0 =
    - \frac{2}{\omega_c}\mel{\Lambda}{d_\epsilon}{\rm CC}^2
    \\ \nonumber
    + 
    2 & \sum_{\mu n} \bra{\Lambda} [d_\epsilon, \tau_\mu (b^\dagger)^n] \ket{\rm CC} X^{d_\epsilon(b+b^\dagger)}_{\mu, n}(0)
    \\ \nonumber
    + & \sum_{\mu \nu n m} F_{\mu n, \nu m} X^{d_\epsilon(b+b^\dagger)}_{\mu n}(0) X^{d_\epsilon(b+b^\dagger)}_{\nu m}(0)
\end{align}
The F-matrix can be simplified to the electronic one,

\begin{align}
    F_{\mu n, \nu m} &= 
    \bra{\Lambda} [[H_e + \omega_c b^\dagger b, \tau_\mu (b^\dagger)^n], \tau_\nu (b^\dagger)^m ] \ket{\rm CC} 
    \nonumber \\
    & = \delta_{nm0} F_{\mu\nu}
\end{align}
and response vector $X^{d_\epsilon(b+b^\dagger)}(0)$ can be related to the response vector $X^{d_\epsilon}$ evaluated at $-\omega_c$,

\begin{align}
    X^{d_\epsilon(b+b^\dagger)}_{\mu n}(0) = &
    \sum_{\nu n} \frac{\bra{\nu, n} e^{-T} d_\epsilon(b+b^\dagger) \ket{\rm CC}}{(\mathbf{A} + n \omega_c \mathbf{I})_{\mu\nu}}
    \\ \nonumber
    = & \delta_{n1}
    \sum_{\nu} \frac{\bra{\nu} e^{-T} d_\epsilon \ket{\rm CC}}{(\mathbf{A} + \omega_c \mathbf{I})_{\mu \nu}}
    \\ \nonumber
    = & \delta_{n1} X^{d_\epsilon}_\mu(-\omega_c)
\end{align}
From this, the electron-photon response function in CC is found to be

\begin{align}
    \llangle d_\epsilon b; d_\epsilon b^\dagger \rrangle_0 &=
    - \frac{1}{\omega_c} \mel{\Lambda}{d_\epsilon}{\rm CC}^2
    \\ \nonumber
    &+
    \sum_{\mu} \bra{\Lambda} [d_\epsilon, \tau_\mu] \ket{\rm CC} X^{d_\epsilon}_{\mu}(-\omega_c).
\end{align}

\bibliography{main}

\end{document}